\documentstyle[12pt,fleqn]{article}
\def\soc{{\rm C}_{60}}
\def\rug{{\rm C}_{70}}
\def\c76{{\rm C}_{76}}
\def\beeq{\begin{equation}}
\def\eneq{\end{equation}}
\def\beeqa{\begin{eqnarray}}
\def\eneqa{\end{eqnarray}}

\addtocounter{section}{-1}
\begin{document}

\begin{center}

\vspace{2in}

{\Large {\bf OPTICAL ABSORPTION IN C$_{\bf {\rm 70}}$ AND C$_{\bf {\rm 76}}$:\\
EFFECTS OF SYMMETRY REDUCTION
}\\
}

\vspace{1cm}

{\rm Kikuo Harigaya\footnote[1]{To whom correspondence should be addressed;
E-mail address: harigaya@etl.go.jp.} and Shuji Abe}\\
\mbox{} \\
{\sl Fundamental Physics Section, Electrotechnical Laboratory,\\
Umezono 1-1-4, Tsukuba 305, Japan}\\

\vspace{1cm}
\end{center}

\noindent
{\Large {\bf Abstract}}\\ %
Optical absorption spectra of $\rug$ and $\c76$ are calculated by a
tight binding model with a long ranged Coulomb interaction.
The spectrum of $\rug$ is in overall agreement
with the experiment of molecules in a solution.
The variations of the spectral shape are discussed relating with the
symmetry reduction from $\rug$ to $\c76$.


\section{Introduction}

Recently, fullerenes C$_N$ with hollow cage structures have been intensively
investigated.  Many optical experiments have been performed, and interesting
properties due to $\pi$ electrons delocalized on molecule surfaces
have been revealed.  They include the optical absorption spectra of
$\soc$ and $\rug$ [1], and the large optical nonlinearity of $\soc$ [2,3].
In order to analyze the optical properties, we have studied the linear
absorption and third harmonic generation of $\soc$ by using a tight
binding model [4] and a model with a long range Coulomb interaction [5].

The purpose of this paper is to extend the calculation of $\soc$ [5] to
higher fullerenes: $\rug$ and $\c76$.  The main purpose is to look at how
the optical absorption changes as the symmetry of the molecule reduces
from $\soc$ and $\rug$ to $\c76$.   We start from a tight binding model
with a long ranged Coulomb interaction.  The Coulomb interaction
is the form of the Ohno potential with the onsite and long range
interaction strengths.  We also consider the broadening of the optical
spectra due to lattice fluctuations by using the bond disorder model.

We shall report calculations in the following way.  In the next section,
our model and its meaning are explained.  Calculation method is also
described shortly.  In \S 3, we report about $\rug$.  The calculated
absorption spectrum is shown, and comparison with experiments in
a solution is made.  We will conclude that there is an overall agreement.
Then, in \S 4, we turn to $\c76$.  We will discuss how the symmetry
reduction appears in the spectrum.  The summary is given in the last
section.

\section{Model and formalism}

We use the following hamiltonian to consider optical spectra of
$\rug$ and $\c76$:
\beeq
H = H_0 + H_{\rm bond} + H_{\rm int}.
\eneq
The first term of Eq. (2.1) is the tight binding model:
\beeq
H_0 = - t \sum_{\langle i,j \rangle, \sigma}
c_{i,\sigma}^\dagger c_{j,\sigma},
\eneq
where $t$ is the hopping integral; $c_{i,\sigma}$ is an annihilation
operator of a $\pi$-electron with spin $\sigma$.
If $t$ depends on the bond length, the results do not change so
strongly, because main contributions come from the strong Coulomb potential.
Effects of zero point vibrations and
thermal fluctuation of the lattice are described
by the bond disorder model which is the second term of Eq. (2.1):
\beeq
H_{\rm bond} = \sum_{\langle i,j \rangle, \sigma} \delta t_{i,j}
c_{i,\sigma}^\dagger c_{j,\sigma}.
\eneq
Here, $\delta t_{i,j}$ is the Gaussian disorder potential at the bond
$\langle i,j \rangle$.   We can estimate the strength of the
disorder (standard deviation) $t_s$ from the results by the
extended Su-Schrieffer-Heeger model [6].  The value would
be $t_s \sim 0.05 - 0.1 t$.  This is of the similar magnitude
as in the fullerene tubules and conjugated polymers.  We shall
treat interactions among $\pi$-electrons by the following model:
\beeqa
H_{\rm int} &=& U \sum_i
(c_{i,\uparrow}^\dagger c_{i,\uparrow} - \frac{1}{2})
(c_{i,\downarrow}^\dagger c_{i,\downarrow} - \frac{1}{2})\\ \nonumber
&+& \sum_{i \neq j} W(r_{i,j})
(\sum_\sigma c_{i,\sigma}^\dagger c_{i,\sigma} - 1)
(\sum_\tau c_{j,\tau}^\dagger c_{j,\tau} - 1),
\eneqa
where $r_{i,j}$ is the distance between the $i$th and $j$th sites and
\beeq
W(r) = \frac{1}{\sqrt{(1/U)^2 + (r/r_0 V)^2}}
\eneq
is the Ohno potential.  The quantity $U$ is the strength of
the onsite interaction, $V$ means the strength of the long range
Coulomb interaction, and $r_0$ is the average bond length.

The model is treated by the Hartree-Fock approximation and the
single excitation configuration interaction method (single CI),
as we used in the previous paper [5].
All the quantities of the energy dimension are shown in the units
of $t$.  Parameters of the Coulomb interaction are changed within
$0 \leq V \leq U \leq 5t$, and we search for the data which reproduce
overall features of experiments of $\soc$ (shown in Ref. [5,7]) and
$\rug$.  The absorption spectra become anisotropic with respect to
the orientation of the molecule against the electric field as
reported in the free electron model (H\"{u}ckel theory) of $\rug$ [8].
We average the anisotropy by rotating the molecule.  The averaged
data can be compared with experiments of molecules in solutions [1].
We also include the lattice fluctuation effects by averaging
over bond disorder potentials as well.  Samples of the orientation
of the molecule and the disorder potential are changed 100 times.
The good convergence is checked by comparing with the results
of the sample number 50.

\section{Optical absorption in C$_{\bf {\rm 70}}$}

Fig. 1(a) shows the molecular structure of the $\rug$.  It has the
rugbyball structure with the $D_{5h}$ symmetry.  The absorption spectrum
is calculated for the Coulomb potential strengths $U = 4t$ and $V = 2t$,
and the bond disorder $t_s = 0.09t$.  The same parameter set has been
successful for $\soc$ when the calculation is compared with the
experiment [7].

Fig. 2(a) shows the calculated optical absorption.  Fig. 2(b) shows
the experimental data of $\rug$ in a solution taken from [1].  The
abscissa is scaled by using $t = 1.9$eV.  The optical gap near $1.0 t$
accords well with the experiment.  The small peaks around $1.8t$
may correspond to those of the experiment in the similar
energy region.   Commonly, there is a dip of the optical
absorption at about $2.0t$ in the calculation and experiment.
There is a maximum at $2.8t$ in the two data.  Therefore, we
conclude that there is an overall agreement between the calculation
and the experiment.  We note that the quantum lattice fluctuation
effects are also simulated well by the bond disorder model.

\section{Optical absorption in C$_{\bf {\rm 76}}$}

There have been two possible structures for $\c76$ which satisfy
the isolated pentagon rule [9].  They have $D_2$ and $T_d$ symmetries.
Owing to the NMR experiment [10], the molecular structure has been identified.
There is only the molecule with $D_2$ symmetry.  The structure is
shown in Fig. 1(b).  We calculate the optical absorption taking
account of the bond disorder $t_s = 0.09t$, and Coulomb interaction
strengths $U = 4t$ and $V = 2t$.  As far as the authors know at present,
there are not detailed experimental data which can be compared with the
calculation.  However, the calculation will be helpful for experimental
developments.

Fig. 3 shows the calculated results.  The average by the rotation
has been performed.  The optical gap is about $0.7t$ and this is
smaller than $1.8t$ of $\soc$ and that of $\rug$.
There are two broad large features around $1.5t$ and $2.5 t$
with fine structures.
We could say that several small peaks around $1.5t$ and $2.6t$ in
the $\rug$ data become fragmented to form the two wide features
in the $\c76$ data.  There is still a dip near $2.0 t$ as in $\rug$.
The position of the dip does not change so much.
These variations could be regarded as owing to the symmetry
reduction from $\soc$ and $\rug$ to $\c76$.

\section{Summary}

We have considered optical spectra of higher fullerenes: $\rug$ and $\c76$.
We have mainly looked at how the optical absorption changes as the symmetry
of the molecule reduces from $\soc$ and $\rug$ to $\c76$.
In the report about $\rug$, we compared the data averaged over the
molecule orientation with experiments in a solution.  We have concluded
that there is an overall agreement about the energy dependence in the
optical spectrum.  In the report of $\c76$, we have discussed how the
symmetry reduction appears in the spectrum.  As we proceed from $\soc$
and $\rug$ to $\c76$, several peaks merge into two large broad peaks
owing to the symmetry reduction.

\mbox{}

\noindent
{\Large {\bf Acknowledgements}}\\
The authors acknowledge helpful correspondence with Dr. Mitsutaka
Fujita.  They also thank Mr. Mitsuho Yoshida for providing them
with molecular coordinates of $\rug$ and $\c76$ which are
generated by the program [11] based on the projection method on
the honeycomb lattice [12].


\mbox{}

\begin{flushleft}
{\Large {\bf References}}
\end{flushleft}

\noindent
$[1]$ Hare, J.P., Kroto, H.W., and Taylor, R., Chem. Phys. Lett.,
1991, \underline{177}, 394.\\
$[2]$ Meth, J.S., Vanherzeele, H., and Wang, Y.,
Chem. Phys. Lett., 1992, \underline{197}, 26.\\
$[3]$ Kafafi, Z.H., Lindle, J.R., Pong, R.G.S., Bartoli, F.J.,
Lingg, L.J., and Milliken, J., Chem. Phys. Lett., 1992,
\underline{188}, 492.\\
$[4]$ Harigaya, K. and Abe, S., Jpn. J. Appl. Phys.,
1992, \underline{31}, L887.\\
$[5]$ Harigaya, K. and Abe, S., J. Lumin., (to be published).\\
$[6]$ Friedman, B. and Harigaya, K., Phys. Rev. B, 1993, \underline{47},
3975; Harigaya, K., Phys. Rev. B, 1993, \underline{48}, 2765.\\
$[7]$ Harigaya, K. and Abe, S., Phys. Rev. B, (preprint).\\
$[8]$ Shumway, J. and Satpathy, S., Chem. Phys. Lett.,
1993, \underline{211}, 595.\\
$[9]$ Manolopoulos, D.E., J. Chem. Soc. Faraday Trans., 1991,
\underline{87}, 2861.\\
$[10]$ Ettl, R., Chao, I., Diederich, F., and Robert, L.W.,
Nature, 1991, \underline{352}, 149.\\
$[11]$ Yoshida, M. and Osawa, E., (preprint).\\
$[12]$ Fujita, M., Saito, R., Dresselhaus, G., and Dresselhaus, M.S.,
Phys. Rev. B, 1992, \underline{45}, 13834.\\

\pagebreak

\begin{flushleft}
{\bf Figure captions}
\end{flushleft}

\noindent
FIG. 1.  Molecular structures for (a) $\rug$ with $D_{5h}$ symmetry
and (b) $\c76$ with $D_2$ symmetry.

{}~

\noindent
FIG. 2.  Optical absorption spectra for $\rug$ shown in the
arbitrary units.  The abscissa is scaled by $t$.
(a) The calculated spectrum with the parameters $U = 4t$
and $V = 2t$.  (b) The experimental spectrum (Ref. [1]) of molecules
in a solution.  We use $t = 1.9$eV.

{}~

\noindent
FIG. 3.  Optical absorption spectrum for $\c76$ shown in the
arbitrary units.  The abscissa is scaled by $t$.
Parameters in the Coulomb potential are $U = 4t$
and $V = 2t$.

\end{document}